\documentclass[5p]{elsarticle}
\date{}
\usepackage{amsmath,amsfonts}
\usepackage{algorithmic}
\usepackage{algorithm}
\usepackage{array}
\usepackage[caption=false,font=normalsize,labelfont=sf,textfont=sf]{subfig}
\usepackage{textcomp}
\usepackage{stfloats}
\usepackage{url}
\usepackage{verbatim}
\usepackage{graphicx}
\usepackage{multirow}
\usepackage{listings}
\usepackage{booktabs}
\usepackage[framemethod=TikZ]{mdframed}
\usepackage{enumerate} 
\usepackage{amssymb}

\usepackage{float}
\usepackage{tabularx}
\usepackage{makecell}
\usepackage{hyperref}
\usepackage{mathtools}
\usepackage{xcolor}
\usepackage{paralist}

\begin{document}

\begin{frontmatter}

\title{Beyond Task Performance: A Metric-Based Analysis of Sequential Cooperation in Heterogeneous Multi-Agent Destructive Foraging}

\ead{amendoza1@us.es; syanes@us.es; dgutierrezreina@us.es; storal@us.es}
\author{Alejandro Mendoza Barrionuevo\corref{corresponding}}
\cortext[corresponding]{Corresponding author}
\author{Samuel Yanes Luis}
\author{Daniel Gutiérrez Reina}
\author{Sergio L. Toral Marín}

\address{Department of Electronic Engineering, University of Sevilla, Av. de Los Descubrimientos s/n, Sevilla, 41003, Spain}

\begin{abstract}
This work addresses the problem of analyzing cooperation in heterogeneous multi-agent systems which operate under partial observability and temporal role dependency, framed within a destructive multi-agent foraging setting. Unlike most previous studies, which focus primarily on algorithmic performance with respect to task completion, this article proposes a systematic set of general-purpose cooperation metrics aimed at characterizing not only efficiency, but also coordination and dependency between teams and agents, fairness, and sensitivity. These metrics are designed to be transferable to different multi-agent sequential domains similar to foraging.
The proposed suite of metrics is structured into three main categories that jointly provide a multilevel characterization of cooperation: primary metrics, inter-team metrics, and intra-team metrics.
They have been validated in a realistic destructive foraging scenario inspired by dynamic aquatic surface cleaning using heterogeneous autonomous vehicles.
It involves two specialized teams with sequential dependencies: one focused on the search of resources, and another on their destruction. 
Several representative approaches have been evaluated, covering both learning-based algorithms and classical heuristic paradigms.
\end{abstract}


\end{frontmatter}

\section{Introduction}
In several real-world environments multi-agent systems enable complex tasks to be tackled through the cooperation of robots or autonomous agents, such as search and rescue missions, cleanup processes, or environmental exploration, among others. This paradigm is particularly useful when agents have different sensory or action capabilities and must coordinate to achieve a common goal.

\begin{figure}[t]
    \centering
    \includegraphics[width=1\linewidth]{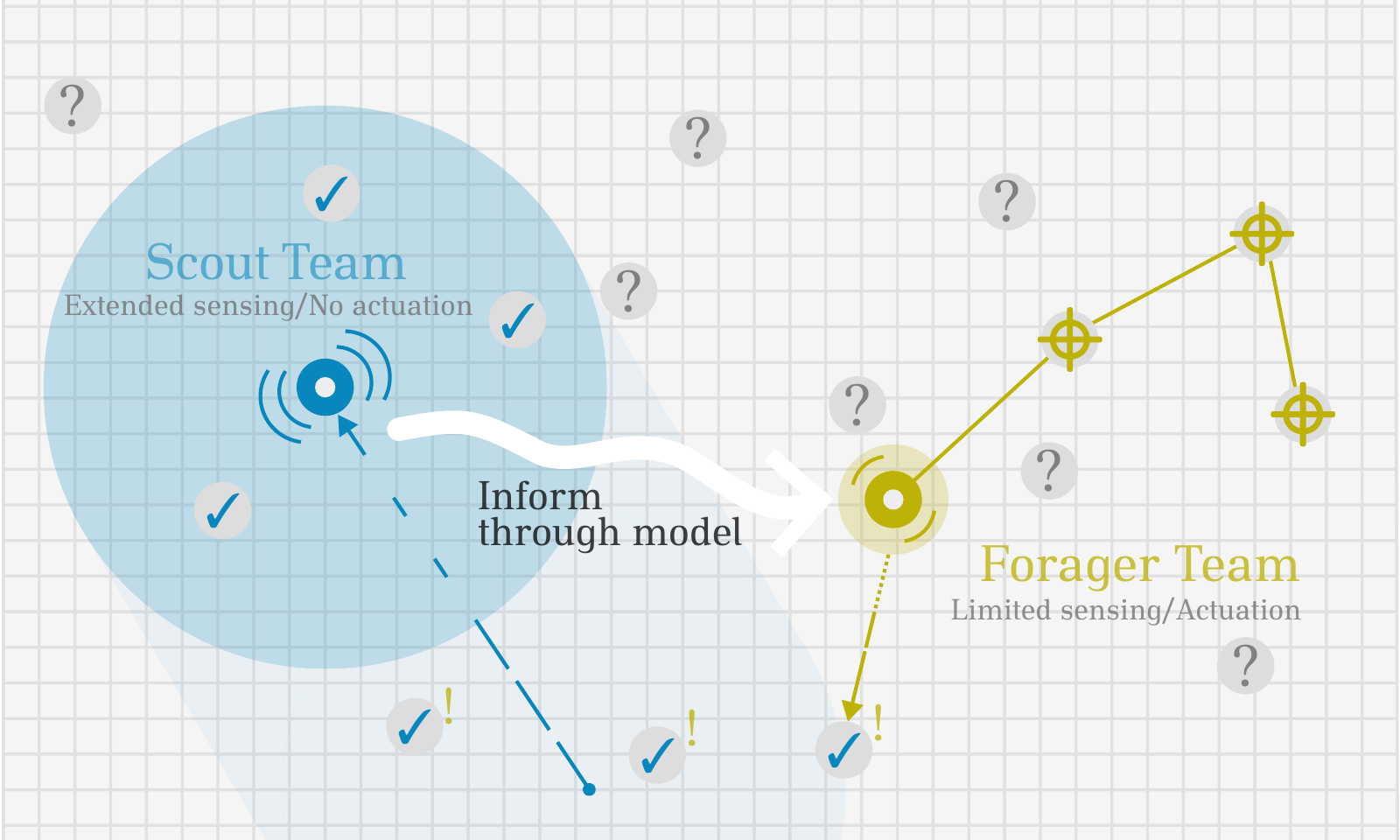}
    \caption{
    A team of agents with extended sensing capabilities explores the environment and discovers resources (scout), while a second team with limited sensing focuses on eliminating the discovered elements (forager).}
    \label{fig:graphical_abstract}
\end{figure}

Many of these scenarios can be described within the general framework of multi-agent foraging (MAF), a problem in which agents must explore the space, locate resources, and act on them, balancing search efficiency and the coordinated use of shared information \cite{state_art_foraging}. It is usually defined as a two-step task known as searching and homing \cite{ostergaard}, but many variations can be derived from this starting point~\cite{state_art_foraging}. In this paper, we focus on what is known as destructive foraging, where resources are removed or consumed when visited \cite{destructive_foraging_bartumeus}. More specifically, it addresses a scenario of sequential cooperation between two heterogeneous teams of agents with different roles: a group of searchers, who are responsible for the search task of foraging, and a group of destroyers, in charge of visiting and removing items.
Beyond the conceptual formulation of the problem, heterogeneity naturally arises in real-world robotic deployments, where teams are rarely composed of identical agents. Instead, robots often exhibit different morphologies, sensing capabilities, mobility constraints, and energy profiles due to design and operational constraints, commonly employing specialized robots optimized for complementary roles.
This results in asymmetric information acquisition and action capabilities across agents, making cooperation not only beneficial but necessary, which reinforces the relevance of studying cooperation under partial observability and role specialization.

Unlike classical approaches to multi-agent foraging, this scenario combines dynamic and destructive resources, partial observability, and temporal dependency between roles, characteristics that have been poorly explored in the literature.
Although the MAF problem has been widely studied in the literature~\cite{state_art_foraging}, most existing approaches focus on developing algorithms that optimize performance with respect to specific task-dependent metrics. However, as far as the authors are aware, no general metrics have been proposed that can be consistently applied across different variants of the problem, nor have any studies been conducted that specifically address quantitative measures of cooperation, whether between heterogeneous teams or among individual agents.

In this work, destructive foraging is applied to a real-world problem: the cleanup of plastic waste floating in aquatic environments using autonomous surface vehicles (ASVs), also known as search-and-clean cooperative problem. In this scenario, agents must operate under dynamic and uncertain conditions, in which waste moves over time due to environmental factors such as wind and currents. To make this study generalizable, the environment is modeled as a graph, where nodes represent spatial locations that may contain one or more waste items. As illustrated in Figure~\ref{fig:graphical_abstract}, scouts are lightweight and fast agents with a wide perception range; they can move several nodes per step and are responsible for updating a shared model of the waste distribution. Foragers, in contrast, are slower and heavier agents with limited perception, capable of removing all items present in a node simply by visiting it.
Cooperation between these two roles is sequential and complementary: scouts improve overall knowledge of the situation by discovering and communicating new waste locations, while foragers use this information to plan their movements and remove the waste. Both types of agents act in a synchronized manner and have a limited number of moves to complete their tasks, due to energy limitations. Therefore, the overall objective is twofold: to maximize the amount of waste removed within a fixed time horizon and to maintain an accurate model of the environment throughout the process.
The objective of this work is to analyze how cooperation emerges and evolves among heterogeneous agents that act with limited and temporally dependent information. To this end, we introduce a set of generic cooperation metrics that quantify the degree of cooperation and synergy between agents, going beyond traditional performance metrics that focus solely on task completion.
Therefore, this work presents three main contributions: 
\begin{itemize}
    \item The formalization of a sequential heterogeneous destructive foraging problem, based on the cooperation of two teams: scouts and foragers. 
    It explicitly captures the temporal dependencies between teams with differentiated roles under partial observability and limited dynamic resources in a grid based scenario.
    \item The definition of a novel and descriptive set of cooperation metrics, aimed at evaluating the efficiency of inter-team and intra-team interaction in heterogeneous multi-agent foraging systems. These metrics address a well-known limitation of existing evaluations in multi-agent foraging, going beyond the final task performance, and allow for a better understanding of the mechanisms that drive cooperative behavior.    
    \item An experimental analysis that studies how different configurations and algorithms affect the overall performance of the system through the proposed metrics.
    They are used to analyze the emergence of coordination patterns, dependencies, robustness to noise, and distribution of effort. It validates the practical utility of the metrics as diagnostic tools, showing that they enable a deeper understanding of why certain approaches perform better, under which conditions cooperation degrades, and how design choices affect collective behavior.
\end{itemize}

\begin{table*}[th]
\centering
\caption{Summary of foraging works and it metrics employed.}
\label{tab:algorithms_comparative}
\scriptsize
\begin{tabular}{cp{4.0cm}p{5cm}p{6.2cm}}
\toprule
\textbf{Ref} 
& \textbf{Foraging Variant} & \textbf{Algorithm} & \textbf{Metrics Employed} \\
\midrule

\cite{bezzo2015exploitingheterogeneousroboticsystems} & Heterogeneous central place MAF & Genetic Algorithm & Average coverage of the environment; Resource collection rate  \\
\midrule

\cite{ARDINY2024104794} & Homogeneous searching MAF & Lévy flight and stigmergy methods & Unexplored area, Exploration time \\
\midrule

\cite{destructive_foraging_bartumeus} & Homogeneous Destructive and non-destructive MAF & Lévy-Modulated Correlated Random Walk & Search efficiency (ratio of visited target sites to total distance traveled); Mean Square Displacement \\
\midrule

\cite{gooreDestructiveForaging} & Homogeneous destructive MAF  &  Linear Reward-Inaction with Lévy Flight distribution & Area covered \\
\midrule

\cite{levywalks_destructiveforaging} & Homogeneous destructive MAF & Lévy Walk with Cauchy distribution & Total number of targets found\\
\midrule

\cite{multiUAVSearchDestroyCoalitions} & Heterogeneous destructive MAF & Optimal Coalition Formation; Random Search; Lanes Based Search; Grid Based Search. & Average mission completion time \\
\midrule

\cite{mutiforaging} & Homogeneous and heterogeneous multi-foraging & Motor Schema-based Reactive Control & Number of items delivered; Diversity metric\\
\midrule

\cite{oflynn2025emergencerolesroboticteams} & Heterogeneous MAF & Deep Q-learning & Average reward; Average resource collection \\
\midrule

\cite{rl_nondestructive} & Single-agent foraging & Reinforcement Learning (Projective Simulation) & Average search efficiency \\
\midrule

\cite{formic} & Homogeneous central place MAF & Asynchronous Advantage Actor-Critic & Total food deposited, Ablation study \\
\midrule

\cite{ZEDADRA2016302} & Homogeneous central place MAF & Cooperative Switching Algorithm for Foraging & Average foraging time, Total food returned, Average path length, Scalability study\\
\midrule

\cite{herdforaging_dualenv} & Homogeneous destructive MAF & Dual-environmental herd-foraging-based coverage path planning & Number of time steps, Number of missing points, Number of covered target points, Average distance\\
\midrule

\cite{radioforaging} & Heterogeneous multi-foraging & Biological foraging inspired communication & Power consumption, End-to-end latency\\
\midrule

\cite{KARPOVA2025105082} & Heterogeneous MAF & Ant-inspired navigation & Successful returns \\
\midrule

\cite{DEBIE2025105133} & Heterogeneous MAF & Task-specialised
swarm (S-swarm) & Resource collection, Exploration coverage, Task Completion Time\\
\midrule

\cite{objectsearchrobots} & Single-agent multi-foraging & Priori knowledge-based approach & Average path length, Time spent, Success rate  \\

\bottomrule
\end{tabular}
\end{table*}

\section{Related Work}\label{Related Work}
Research on MAF and cooperative exploration has grown substantially over the past decades, especially due to advances in robotics and algorithms, as seen in \cite{state_art_foraging}. 
As robotic systems evolved, principles of foraging were progressively adapted to engineered multi-agent teams, leading to a variety of approaches for coordinated search, resource collection, and task allocation. 
More recent work has focused on heterogeneous teams, where agents possess different sensing, mobility, or actuation capabilities, such as  \cite{bezzo2015exploitingheterogeneousroboticsystems}.
There are many variants of foraging, depending on different characteristics such as whether the type of items being collected is the same or different (multi-foraging) \cite{mutiforaging}, whether homing can be done to a single point or there are several \cite{debout_multilocation}, or whether they are directly destroyed when found, as in destructive foraging \cite{destructive_foraging_bartumeus}.
Table~\ref{tab:algorithms_comparative} shows a summary of the algorithms mentioned in this section.

Work \cite{oflynn2025emergencerolesroboticteams} presents a centralized reinforcement learning (RL) strategy designed for a competitive multi-agent foraging scenario. In their formulation, the objective is to maximize the collection of available resources while simultaneously competing against an adversarial team that pursues the same objective. Contrary to our work, there are conflicting interests between teams, but the collection task they propose can be classified as destructive foraging. 
In \cite{rl_nondestructive} authors model a forager as RL single agent in a non-destructive search environment with immobile and scattered targets. The agent decides to continue or make a random turn based on the perceived state. These two studies base their evaluation metrics only on the resources collected. In \cite{DEBIE2025105133} a  heterogeneous robotic swarm approach based on role specialization is proposed to optimize MAF. Similar to our work, the mission is divided in into two interdependent subtasks performed by different teams: searching and transportation. Although they use more performance evaluation metrics than previous studies, the authors again do not use metrics to analyze cooperation.

The problem of central place foraging studied in \cite{formic} is addressed using the asynchronous advantage actor-critic (A3C) algorithm. The system is based on stigmergy, allowing agents to communicate using virtual pheromones. Although it conducts an ablation study, the only metric focused on the behavior of agents in relation to foraging is based on total food deposited, without studying other collaboration metrics. A similar case study is addressed in \cite{ZEDADRA2016302}, which proposes solving the problem in large-scale systems (with hundreds or thousands of simple agents) with an algorithm called Cooperative Switching Algorithm for Foraging, based on swarm intelligence. The metrics employed evaluate the average foraging time required to complete the mission, the total amount of food returned to the nest, the average length of the optimized route, and the scalability of the system. Work \cite{ARDINY2024104794} investigates the application of bio-inspired swarm robotics for the exploration and mapping of hazardous radioactive environments, such as nuclear facilities. It is, therefore, a search-only foraging strategy. The success of the system is primarily measured by exploration time and percentage of environment coverage.

Work \cite{KARPOVA2025105082} describes a robotic navigation algorithm based on the biological behavior of Formica rufa ants during their search for food. The method stands out for being a map-free strategy, where robots use visual landmarks, compass data, and elapsed time to memorize and repeat routes. In the same way as our work, agents are divided into the roles of scout and forager. However, the evaluation metrics are very limited, with little analysis of success in terms of return.
In \cite{herdforaging_dualenv}, the main task of the agents is the path planning for areas that need to be covered, within which there are target points that must be visited. These points disappear once visited, so it can be classified as a destructive foraging problem. The metrics used focus exclusively on the number of steps it takes agents to visit target points, or the number of points they visit.
In \cite{radioforaging}, the prey model of optimal foraging theory is used, treating nodes as foragers that autonomously decide which communication channels to “capture” based on their energy and time efficiency.
Other articles do not explicitly refer to the problem as foraging, but they are analogous to it, allowing the behavior of agents to be modeled as a strategic search for resources. For example, in the service robotics described in \cite{objectsearchrobots}, the search for dynamic objects in the home follows a foraging logic.

In \cite{levywalks_destructiveforaging}, a search strategy based on Lévy Walks is proposed and applied to a destructive foraging scenario in the field of swarm robotics. The novelty of the proposed algorithm lies in increasing the time spent on exploitation (intensive search) each time a target is located.
Other works, such as \cite{gooreDestructiveForaging}, have explored destructive foraging proposing decentralized agent-based strategies for searching and eliminating randomly distributed targets.
In \cite{multiUAVSearchDestroyCoalitions}, search-and-destroy missions are investigated in multi unmanned aerial vehicle systems, analysing coordinated exploration strategies and coalition-formation mechanisms for resource-dependent target elimination. Diverging from our research, this work does not address sequential cooperation between specialised teams, and the evaluation focuses solely on mission completion time as the main performance indicator, without considering agents cooperation-related metrics.

Related work reflects a broader trend across the literature: although many works address closely related multi-agent foraging or search-and-elimination problems, there is little consistency in the evaluation methodologies, as seen in Table~\ref{tab:algorithms_comparative}. Each study tends to define its own task-specific performance metrics, typically oriented toward final outcomes (e.g., total resources retrieved or time to mission completion), which makes approaches difficult to compare. Moreover, even when equivalent metrics are employed, they are often insufficient to reveal the cooperative mechanisms and interaction patterns that give rise to the observed performance.
This limitation becomes particularly critical when applying black-box optimization techniques, such as deep reinforcement learning or metaheuristic methods. In these cases, the interpretation of the resulting behaviors is largely delegated to a single utility function (e.g., reward or fitness), which prevents us from knowing how cooperation, redundancy, or role specialization actually arise.
The literature still lacks a unified framework of general-purpose metrics suitable for evaluating heterogeneous cooperative foraging scenarios, as it tends to propose domain-specific measures.
To address this gap, this paper proposes the study of a complementary research direction focused on the formalization of evaluation metrics and benchmarking methodologies for cooperative multi-agent systems.
The proposed metrics aim to characterize aspects such as coverage efficiency, redundancy, fairness, cooperation strength, and robustness to failures, enabling a more interpretable and systematic analysis of policies beyond raw task performance.

\section{Problem Formulation and Representation}

This section describes a multi-agent scenario that will be used later to illustrate and evaluate the proposed cooperation metrics. The objective of this formulation is to provide a general and sufficiently abstract framework that allows for the analysis of sequential cooperation dynamics in heterogeneous teams under partial observability and dynamic resources. 
The definition of the environment, the agents, and their capabilities will serve as an experimental basis for studying how different coordination strategies affect collective performance beyond traditional metrics of final results.

\subsection{Foraging Environment}
The problem considered in this work consists of a sequential cooperative destructive foraging scenario involving two heterogeneous teams of agents operating in a dynamic and partially observable environment. 
The agents must jointly explore and clean an area of size $H \times W$ represented as a graph $G = (V,E)$, where each node $v_{i,j} \in V$ can contain one or more waste items. 
The set of edges \( E \subseteq V \times V \) defines the connectivity between adjacent nodes, specifying all feasible transitions that an agent can perform between neighboring positions, which represents the possible movements between two adjacent nodes. In this work each node can be linked to up to eight surrounding nodes. Therefore, nodes with fewer than eight edges correspond to positions adjacent to obstacles or boundaries in the environment, which are supposed to be known. Agents can take actions $a \in A$ to move following one of the available edges.

The team of scouts is made up of lightweight and fast agents with a wide sensing range, capable of moving two nodes per time step and updating the shared model about the waste distribution. The team of foragers, is made up of heavier and slower agents (one node per time step) with a limited perception range. When a forager agent visits a node, all waste elements located there are removed, modeling a destructive resource interaction.

Both types of agents move synchronously and during each episode the agents are subject to a maximum number of time steps $T$. Each agent has only partial knowledge of the environment, obtained from a shared model.
The overall objective is twofold: first, to maximize the number of waste items removed within the time horizon $T$, and second, to minimize the discrepancy between the estimated and actual waste distributions, expressed as a modeling error $E(t)$, typically computed through the root mean square error (RMSE) between the real and perceived states of the environment.
This formulation aligns with recent perspectives in multi-agent foraging, where the efficiency of resource collection is increasingly linked to the acquisition of information~\cite{pitonakova2016information, pitonakova2018information}. These studies highlight that the collective performance of the mission depends on how accurately agents perceive and share environmental information. 

The positions of the fleet of $N$ vehicles is formulated as a set $\mathcal{P} = \{ p_n \mid n = 1, 2, \ldots, N \}$, where $p_n$ is the position of the vehicle $n$. Each of the possible positions a vehicle can take is within a node, such as $p_n = v_{i,j} \in V$. The graph is represented as a matrix $M[i, j] \in \{0, 1\}$ of size $H\times W$, where $M[i, j] = 1$ corresponds to a navigable node $v_{i,j}$, and 0 otherwise.

\begin{figure}[t]
    \centering
    \includegraphics[width=1\linewidth]{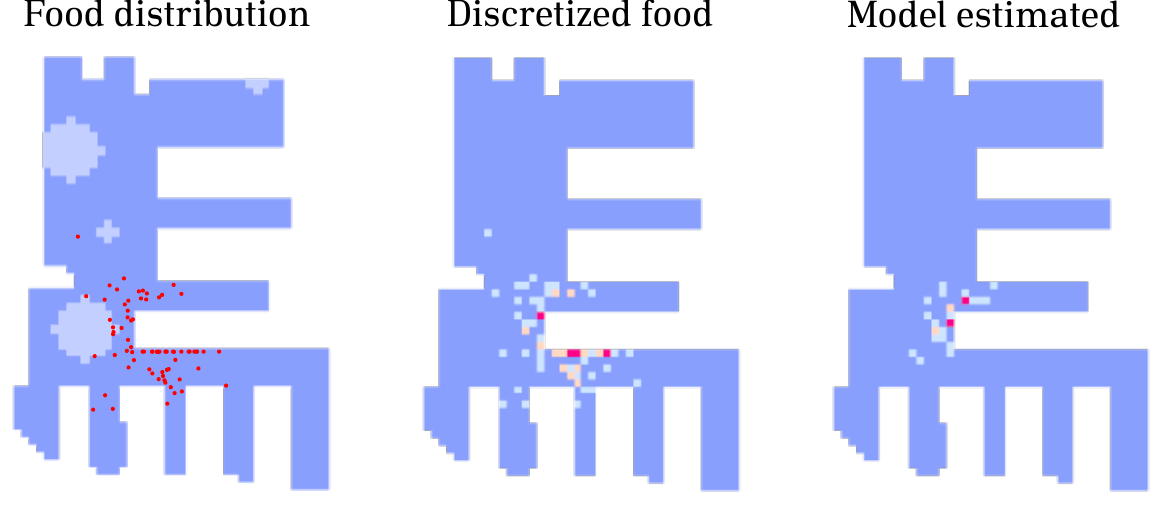}
    \caption{Scenario of the foraging environment with food distribution and differentiated field of vision of the scout and forager agents (left), discretized food distribution (middle), and model estimated based on the perceptions of the agents (right).}
    \label{fig:food_render}
\end{figure}

\subsection{Items Dynamics}
The set of trash items, showed in Figure~\ref{fig:food_render}, is represented as \( B = \{ b_k = (x_k, y_k) \mid k = 1, 2, \ldots, K \} \), where each element \( b_k \) specifies the spatial coordinates of a single item within a continuous reference frame. To follow the nomenclature of foraging, the trash items will be referred to as \emph{food}.
The total number of items, \( K \), is taken from a normal distribution at the beginning of each episode. A random node among the visitable positions is chosen as the contamination source, around which the elements of \( B \) are generated following a multivariate normal distribution. Consequently, each episode starts with a distinct spatial configuration of food, and no additional items are introduced during the foraging process.
The positions of items evolve dynamically throughout the episode to emulate environmental influences. Specifically, two components are considered: wind and random perturbations. Wind is modeled as a constant velocity vector \( v_{\text{wind}} \) that applies equally to all items, sampled from a uniform distribution at the beginning of the episode. Random fluctuations are introduced through individual velocity variations \( v_{\text{rand}} \), sampled independently from uniform distributions at each time step. Thus, the position of each item is updated as follows:
\[
b_{k}^{t+1} = b_{k}^{t} + \Delta t \left( w_{\text{wind}} \cdot v_{\text{wind}} + w_{\text{rand}} \cdot v_{\text{rand}} \right),
\]
where \( w_{\text{wind}} \) and \( w_{\text{rand}} \) are weighting coefficients, both set to 1 in this work.

The spatial distribution of food can be represented as a matrix \( Y \) of dimensions \( H \times W \), 
where each element \( Y[i,j] \) corresponds to the number of items from the set \( B \) at a given time step that fall within the  physical region $\text{Area}(i,j)$ associated with node \( v_{i,j} \). 
Formally, this can be expressed as
\[
Y[i,j] = \left| \left\{ b_k \in B \mid (x_k, y_k) \in \text{Area}(i,j) \right\} \right|,
\]
Hence, \( Y[i,j] = 0 \) indicates that the cell is free of resources, 
while a positive value of \( Y[i,j] \) denotes the presence of one or more food items.

Whenever a forager agent visits a cell, all food resources located within that cell are considered collected and removed from the set \( B \). Forager agents are assumed to have unlimited storage capacity, and the pickup process is managed internally by the agent’s local control system, 
without influencing the decision-making logic of the foraging. For that reason, the foraging environment is considered destructive.
The quantity of food collected by a forager positioned at \( p_n \in \mathcal{P} \) at time \( t \) 
can therefore be defined as
\[
C(p_n, Y) = Y[i,j] \quad \text{such that} \quad v_{i,j} = p_n.
\]

The range of visual coverage \( \Theta(p_n) \) denote the set of nodes within the field of view of a vehicle located at position \( p_n \), defined as all nodes contained within a sensing radius \( \rho \):
\[
\Theta(p_n) = \{\, v \in V \mid \| v - p_n \| < \rho \,\}.
\]
In this work, the sensing radius of scouts is set to (\( \rho = 4 \)), while the sensing radius of foragers is restricted to a single-node, as they have limited vision. This can be seen on the left of Figure~\ref{fig:food_render}, where the fields of vision of the agents show which portions of the map they can observe, with the scouts having a much broader one.
The estimated food distribution is shared among all agents and represented by a matrix 
\( \hat{Y}[i,j] \), initially set to zero. Based on these constraints, agents update at each time step the model based on what they see (on the right of Figure~\ref{fig:food_render}), i.e., each cell value in \( \hat{Y} \) is updated to match the true value of \( Y[i,j] \) whenever the corresponding node falls within the field of view of any agent:
\[
\hat{Y}[i,j] \leftarrow Y[i,j] \quad \text{if} \quad v_{i,j} \in \Theta(p_n).
\]

\section{Metrics Definition}
As previously discussed in Section~\ref{Related Work}, although multi-agent foraging has been extensively explored \cite{state_art_foraging}, the evaluation metrics employed in most studies are not easily generalizable, as they are closely coupled to the specific task or environment. 
Furthermore, most of them focus exclusively on performance-based evaluation criteria, and, to the best of the authors' knowledge, there is still a lack of general-purpose metrics capable of consistently evaluating cooperation between different variants of the multi-agent foraging problem.
To address these limitations, this section introduces a set of generalized metrics aimed at evaluating not only overall system performance but also the cooperative interactions that emerge between agents and teams during sequential foraging tasks. 
These metrics are organized into several complementary categories:

\begin{enumerate}
    \item \textbf{Primary metrics}, which evaluate the overall performance of the multi-agent system in terms of resource removal or accuracy of information.
    \item \textbf{Inter-team metrics}, which capture the cooperation between heterogeneous teams in terms of dependency and coordination.
    \item \textbf{Intra-team metrics}, which quantify contributions and fairness within each agent of same the team.
\end{enumerate}

\subsection{Primary metrics}
These metrics correspond to the class of performance indicators most commonly used to evaluate multi-agent algorithms, as they summarize overall task achievement and efficiency at the system level. While these metrics do not explicitly capture cooperative behaviour or agent interactions, they provide essential context for interpreting more detailed the subsequent inter and intra-team analyses. In this work, we extend and systematize this set of primary metrics to establish a common baseline that supports consistent comparison across algorithms.

\paragraph{Percentage of Target Achieved (PTA)}
This metric measures the overall effectiveness of the system in completing a main objective.  
Let \(U\) denote a utility variable associated with task completion (e.g., number of resources discovered or removed), and let \(U^{UB}\) denote its corresponding upper bound.  
In the destructive foraging context presented, for foragers it expresses the proportion of total resources successfully removed from the environment during an episode relative to the total number of items initially present, while for scouts it expresses the proportion of total resources discovered.  
It is defined as:
\begin{equation}
\begin{aligned}
    PTA = \frac{U}{U^{UB}}\cdot100 \in [0,100],
\end{aligned}
\end{equation}
A $PTA$ value equal to~100 means that agents accomplished the task completely, while lower values indicate partial or inefficient completion.  
This provides a simple global indicator of system performance, independent of the internal coordination or individual contributions of the agents.

This is a metric commonly employed in prior work, as discussed in Section~\ref{Related Work} in works such as~\cite{bezzo2015exploitingheterogeneousroboticsystems,levywalks_destructiveforaging, oflynn2025emergencerolesroboticteams}, although under different names or slightly varying formulations. 
In this work, PTA is introduced as a unified metric that can be applied to quantify the degree of task completion in a wide range of foraging scenarios, provided that the total number of targets or objectives is known or can be estimated. 
It is a externally observable metric, as it relies on ground-truth information about task completion, and does not require access to the internal states or policies of the agents. As such, it can be computed either online, during execution, or offline, once an episode has concluded.

\paragraph{Root Mean Squared Error (RMSE)}
In the context of foraging, this metric quantifies the discrepancy between the estimated map $\hat Y$ of food distribution and the real underlying distribution $Y$. It measures the accuracy of the model and reflects how effectively the agents maintain an up-to-date representation of the environment. As in this work $Y$ and $\hat Y$ are sparse matrices and the elementwise error will produce high error variations, a Gaussian Filter $G$ with $\sigma = 1$ is applied to compare the density of the food distribution:
\begin{equation}
    { RMSE(t) = \sqrt{  \frac{1}{|V|} \sum_{v \in V}\left(G \circledast Y_{v,t} - G \circledast  \hat Y_{v,t}\right)^2} }
\end{equation}

Lower values indicate a more accurate and consistent model estimate across nodes, whereas higher values imply mismatches due to outdated, incomplete, or incorrect information.

Although RMSE has not been explicitly employed in the related works discussed in Section~\ref{Related Work}, it is a widely adopted metric in problems involving the estimation of a ground truth state (e.g.,~\cite{popovic2020informative, mendozab}), whenever it is available.
Its formulation is independent of the specific task semantics, making it applicable to any foraging scenario that involves discovery, tracking, or estimation of an underlying ground truth, whether static or dynamic, and it can be computed online or offline.

\paragraph{Normalized Time to \(X\) ($NT_X$)}
This metric assesses temporal efficiency, i.e., how quickly the system reaches a given percentage \(X\) of the total objective, expressed in a normalized scale to allow comparison across different domains or episode durations.  
It measures the proportion of the total available time required to achieve the target, and is defined as:

\begin{equation}
    NT_X = \frac{t_X}{T} \in [0,1],
\end{equation}

where \(t_X\) is the time step at which the system first reaches a percentage \(X\) of the total target (for example, \(PTA_t \geq X\)), and \(T\) is the maximum possible number of time steps for each the episode.  

A value of \(T_X = 0\) indicates that the goal fraction \(X\) was achieved immediately at the beginning of the episode, while \(T_X = 1\) means that it was never reached within the allowed time.  
This normalized formulation provides a dimensionless indicator of temporal efficiency, enabling direct comparison between experiments with different time horizons or environmental dynamics.

Although this metric has not been employed in this exact form in prior works seen in Section~\ref{Related Work}, similar time-to-completion metrics are commonly used in the literature to assess efficiency, as in \cite{multiUAVSearchDestroyCoalitions} or \cite{ZEDADRA2016302}, although often in unnormalized or task-specific forms.
In the same way as PTA metric, it is a externally observable metric, as it relies on ground-truth information about task completion, and it can be computed either online, as soon as the threshold \(X\) is reached, or offline.

\paragraph{Throughput}
It provides a normalized indicator that allows comparison between configurations involving different numbers of agents or episode durations. 
Formally, it can be defined as:

\begin{equation}
    Throughput = \frac{Resources \: Collected}{|N_f| \cdot T},
\end{equation}

where $Resources \: Collected$ denotes the total number of resources handled by forager agents during the episode, 
$|N_f|$ represents the total number of foragers, and $T$ is the number of time steps. 
This metric could also be used to calculate the throughput of the scout team in its search for items, changing resources collected to discovered.

While similar notions of efficiency appear implicitly in prior foraging and search-and-service studies \cite{destructive_foraging_bartumeus, rl_nondestructive}, they are often reported in episode-level form rather than as a normalized per-step and per-agent metric.
Although this metric can be computed both online and offline, it is preferable to evaluate it at the end of the episode, as intermediate values may be strongly influenced by transient effects.
Due to its normalized formulation, throughput can be applied to foraging scenarios defined over a temporal evolution, whether expressed in continuous time or discrete steps, as long as they involve resource collection, servicing, or elimination, regardless of the specific task semantics or scale of the environment.

\paragraph{Idleness Metrics}
This metrics evaluate how effectively the agents contribute to reducing the idleness of the environment over time, that is, the duration since each region or node was last visited by an agent. 

Formally, the idleness of a node \( v_{i,j} \) at time \( t \), denoted as \( I_{i,j}(t) \in [0,1] \), represents the normalized measure of how long the node has remained unvisited. 
Idleness increases progressively at each time step according to a forgetting factor \( f \in (0,1) \), which controls the rate at which information about previously visited nodes becomes outdated. 
Whenever a node is visited, its idleness value is reset to zero. 
Thus, nodes that are frequently revisited maintain low idleness values, while those ignored for longer periods gradually approach to 1.

The \textit{Mean Idleness} (MI) of the environment can then be defined as the average idleness across all visitable nodes \( |V| \):

\begin{equation}
    MI(t) = \frac{1}{|V|} \sum_{v_{i,j} \in V} I_{i,j}(t),
\end{equation}

The \textit{Idleness Reduction Rate} (IRR) measures the rate at which this average idleness decreases over time, capturing the collective effort of the agents to keep the environment well explored and updated:

\begin{equation}
    IRR(t) = -\frac{dI(t)}{dt} \approx \frac{I(t) - I(t+1)}{\Delta t}.
\end{equation}

Higher values of $IRR(t)$ indicate that agents are efficiently visiting unexplored or outdated regions, thereby reducing the global idleness and improving environmental awareness, maintaining spatial freshness.
Although it has not been explicitly employed in the foraging literature reviewed in Section~\ref{Related Work}, it is commonly used in related domains such as robot exploration and patrolling problems~\cite{yanes_ypacarai, idleness_patrol}.
This metric is fully observable whenever visit timestamps are available and can be computed both online, to monitor exploration dynamics in real time, and offline, by aggregating statistics over a completed episode. Idleness metrics are particularly useful in foraging scenarios that require food search phases, especially if food appears throughout the episode and requires balanced revisits.

\subsection{Inter-team metrics}
Inter-team metrics are designed to evaluate how effectively heterogeneous teams with different roles coordinate and depend on each other. In sequential cooperative settings, the performance of teams are tightly coupled to the actions performed by the others, making isolated evaluation insufficient. The metrics presented in this section capture key aspects of this interaction, including temporal alignment, cooperative dependency, or robustness to degradation, among others.

\paragraph{Discovery-to-Service Latency (DSL)}
This metric captures the temporal dependency between sequential roles in a cooperative foraging system, measuring the efficiency of inter-team coordination. 
Formally, DSL is defined as the time gap between the moment a resource is detected by a scout agent, \( t_d \), and the moment it is effectively processed or removed by a forager agent, \( t_c \), normalized over the total number of time steps $T$:

\begin{equation}
    DSL = \frac{t_c - t_d}{T}.
\end{equation}

Lower values of \( DSL \) indicate a faster and more efficient cooperative response, reflecting strong temporal coupling between tasks.

\paragraph{Inter-Team Temporal Lag (ITL)}
This metric quantifies the temporal misalignment between the peak activity or maximum performance of two cooperating teams. It captures how synchronized their contributions are over the course of the episode.
Let \(R_1(t)\) and \(R_2(t)\) denote the time series of a performance metric (e.g., PTA) for teams 1 and 2, respectively, over a total of \(T\) time steps.  
Define \(t^{*}_1 = \arg\max_t R_1(t)\) and \(t^{*}_2 = \arg\max_t R_2(t)\) as the time steps at which each team reaches its peak performance.

The inter-team temporal lag is then given by:
\begin{equation}
ITL = \frac{|t^{*}_1 - t^{*}_2|}{T} \in [0,1],
\end{equation}
where \(T\) is the total duration of the episode.

Thus, \(ITL = 0\) indicates perfect temporal synchronization between teams, i.e., their peak activity occurs simultaneously. Higher values of \(ITL\) indicate increasing temporal desynchronization, meaning that one team reaches its performance peak significantly earlier or later than the other.
It is important to note that in sequential or role-specialized tasks, such as search followed by destruction, a certain temporal offset between teams is expected and even desirable. In these cases, ITL reflects the coordination structure of the task rather than inefficiency. Metrics should always be interpreted in context, not in isolation.

\paragraph{Cooperative Success Ratio (CSR)}
This metric quantifies the ratio in which the successful completion of the foraging task depended on cooperation between teams with complementary functions. In present environment, it is the number of items collected by the forager team that were first discovered by the scout team.
It can be expressed as:

\begin{equation}
    CSR = \frac{\text{Tasks performed through cooperation}}{\text{Total tasks}}.
\end{equation}
A high value of \( CSR \) indicates that most foraging tasks benefit from interdependent coordination between teams.

\paragraph{Cooperation Sensitivity under Stochastic Corruption (CSSC)}
We introduce a controlled corruption experiment to quantify how the performance of a target team (denoted \(T_A\)) depends on the behavioral quality of a partner team (denoted \(T_B\)). Team \(T_B\) executes a stochasticized policy \(\pi_{T_B}^\varepsilon\) that, at each decision step, selects with probability \(1-\varepsilon\) the action $a$ prescribed by its nominal policy \(\pi_{T_B}\) and with probability \(\varepsilon\) a uniformly random action from its action set. The corruption parameter 
\(\varepsilon \in [0,1]\) interpolates between perfect behavior (\(\varepsilon=0\)) and fully random behavior (\(\varepsilon=1\)).

Formally, for each \(\varepsilon\) the mixed policy for an agent \(t_b\in T_B\) is
\begin{equation}
\pi_{t_b}^{\varepsilon}(a\mid s) = (1-\varepsilon)\,\pi_{t_b}(a\mid s) + \varepsilon\,\frac{1}{|A_{t_b}|},
\end{equation}
where \(A_{t_b}\) is the discrete action set of agent \({t_b}\).

The degradation curve is obtained by progressively injecting stochastic noise into the action-selection policy of team \(T_B\) through the corruption parameter \(\varepsilon_{T_B} \in [0,1]\), and measuring the resulting performance of team \(T_A\). For each value of \(\varepsilon_{T_B} \), a corresponding performance value \(R_{T_A}(\varepsilon_{T_B} )\) is recorded, yielding the degradation curve \(\varepsilon_{T_B}  \mapsto R_{T_A}(\varepsilon_{T_B} )\). This curve characterizes how sensitive the performance of team \(T_A\) is to the reliability of team \(T_B\).

The relationship between \(R_{T_A}(\varepsilon_{T_B} )\) and the degradation factor \(\varepsilon_{T_B} \) is then modeled by means of linear regression. By computing the slope of the best-fitting line, a scalar sensitivity indicator is obtained, named \emph{Sensitivity Slope}, which quantifies how strongly the performance of team \(T_A\) deteriorates as the behavior of team \(T_B\) becomes increasingly corrupted. The resulting metric captures the degree of inter-team dependence and the robustness of the cooperative process, defined as:

\begin{equation}
SS_{T_A\leftarrow T_B} \;=\; \text{slope}\Big(\text{linreg}\big(\varepsilon_{T_B} , R_{T_A}(\varepsilon_{T_B} )\big)\Big),
\end{equation}

A higher value of \(SS_{T_A}\) reflects a stronger degradation of performance as the partner team becomes less reliable, indicating higher dependence and reduced robustness of cooperation. 
It should be noted that the degradation curve may not always show a single linear approximation. In practice, the curve can present different best-fit slopes for certain sections, corresponding to different sensitivity phases of the cooperative process. For example, one may observe a slight initial degradation for small values of $\epsilon$, followed by a more pronounced decline when it reaches a critical threshold of corruption.

\subsection{Intra-team metrics}
Intra-team metrics focus on the internal coordination dynamics within each team, assessing how responsibilities, effort, and spatial coverage are distributed among agents sharing the same role. These metrics capture aspects such as fairness, redundancy, and collective efficiency, revealing whether team performance emerges from balanced cooperation or from uneven contributions.

\paragraph{Gini Index of Contributions}
This metric evaluates the fairness of workload distribution among agents by measuring inequality in individual contributions to the collective objective. 
It is based on the Gini coefficient \cite{gastwirth1972estimation}, a well-known statistical measure of inequality that ranges from \( 0 \) (perfect equality) to \( 1 \) (maximum inequality). 
In the context of multi-agent foraging, it quantifies how uniformly resources are discovered or collected across the teams. 
Let \( c_n \) denote the number of resources handled by agent \( n \) belonging to $T$ team, with \( n = 1, 2, \ldots, |N_T| \). 
The Gini index of contributions is defined as:

\begin{equation}
    G = \frac{\sum_{i=1}^{|N_T|} \sum_{j=1}^{|N_T|} |c_i - c_j|}{2 \, |N_T|^2 \, \bar{c}},
\end{equation}

where \( \bar{c} \) represents the mean contribution per agent.

A value of \( G = 0 \) indicates perfect equality among agents, meaning that all agents contributed equally to the collective objective. 
The maximum possible value depends on the number of agents on the team \( |N^T| \). 
For the extreme case where a single agent performs all the work while the others contribute nothing, the Gini coefficient reaches $G_{\max} = (|N^T| - 1) / |N^T|$.
This metric provides a fairness-oriented perspective that complements efficiency metrics, allowing the analysis of whether cooperative performance is achieved through balanced or unbalanced contributions.

\paragraph{Coverage Overlap (CO)} 
This metric quantifies, at each time step, the proportion of the currently covered area that is redundantly covered by multiple agents. It reflects the efficiency of spatial coordination: a low value indicates well-distributed coverage, while a high value suggests significant redundancy.

\begin{figure}[t]
    \centering
    \includegraphics[width=0.45\linewidth]{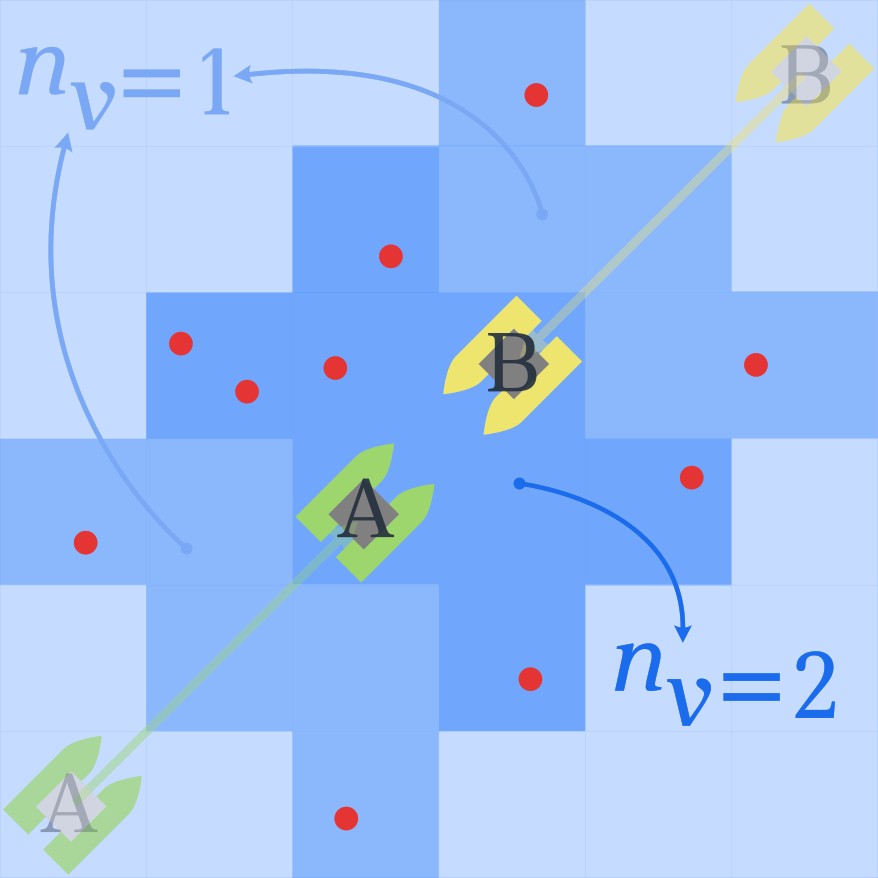}
    \caption{Example of coverage overlap between two agents that share the same nodes in their field of view.}
    \label{fig:overlap}
\end{figure}

Let $\Theta(p_n, t)$ denote the subset of nodes within the field of view of a vehicle located at position $p_n$ at time $t$. For each node \(v \in V\), the number of distinct agents $n_v$ covering the node at time $t$ is defined:
\begin{equation}
n_{v}(t) = \sum_{n \in N} \mathbf{1}\{ v \in \Theta(p_n,t) \},
\end{equation}

Then, the overlap can be expressed as:
\begin{equation}
CO(t) = 
\frac{\sum_{v \in V} \mathbf{1}\{ n_v(t) \ge 2 \}}
{\sum_{v \in V} \mathbf{1}\{ n_v(t) \ge 1 \}}.
\end{equation}

This is graphically illustrated in Figure~\ref{fig:overlap}. Thus, $CO(t) = 0$ represents perfectly non-overlapping coverage, while higher values of $CO(t)$ indicate redundant exploration among agents.

\paragraph{Marginal Contribution per Agent (MC)}
This metric quantifies the individual cooperative impact of each agent in each role by measuring how much the overall team performance decreases when that agent is removed from its team. 

Let \(R_{\text{full}}\) denote a performance metric (e.g., total reward, resources collected, or reduction in expected service cost) achieved by the entire team, and \(R_{-i}\) the performance obtained when agent \(i\) is removed from the team.  
The marginal contribution of agent \(i\) is then defined as:

\begin{equation}
MC_i =
\begin{cases}
\dfrac{R_{\text{full}} - R_{-i}}{R_{\text{full}}} & \text{if greater values of $R$ are better},\\[4mm]
\dfrac{R_{-i} - R_{\text{full}}}{R_{-i}} & \text{if lower values of $R$ are better}.
\end{cases}
\end{equation}

Values close to 0 indicate that the agent has little impact on the mission outcome (high redundancy or limited usefulness), whereas values closer to 1 reveal that the agent plays a critical role. 
If it is negative, this indicates that the system actually performs better when the agent is removed (the agent was detrimental).
This metric also reflects the resilience of the remaining agents, since a small degradation after removal suggests strong compensatory capability within the team.

\section{Experimental Study}

This section aims to analyze how different path planning algorithms impact performance, information exchange, and cooperative efficiency in the foraging environment under study. This environment involves heterogeneous teams operating under partial observability and temporal dependencies.
For this purpose, a study of the variations in the proposed cooperation and performance metrics is conducted, observing how each algorithm influences the emergence of coordination patterns, information flow, and collective efficiency throughout the foraging process. 
Special attention is given to the evolution of inter-team and intra-team metrics, which reveal not only the overall performance but also the underlying cooperative dynamics that drive it.

Several algorithms have been employed for comparative analysis, including both modern learning-based methods and classical heuristic approaches.  
First, a Double Deep Q-Learning (DDQL) strategy is considered, following the architecture and training procedure described in \cite{mendozab}, with similar reward function. This deep reinforcement learning (DRL) method is implemented using shared neural policies per role: all agents belonging to the same team type (e.g., scouts or foragers) query a single neural network that outputs their actions from their local observations. This corresponds to a parameter-sharing scheme with decentralized execution. 
In addition, two non-learning baselines are evaluated. The first is a Greedy strategy, in which agents chooses the action that maximizes instant reward, without further consideration of future consequences.
The second is a Lévy-Walk–based approach, where scout agents perform stochastic exploration using Lévy walks distributions inspired by biological foraging efficiency, while foragers employ shortest-path navigation (Dijkstra) to the closest known resource location. When there are no known food items, foragers follow the same exploratory pattern as scouts.
All algorithms are evaluated within the same foraging environment in an average of the same 100 episodes, sharing the same seed so that food dynamics and appearance have identical variability. The code employed is available on Github (\url{https://github.com/amendb/CollaborationMetrics}) so results can be reproduced.

\begin{figure}[t]
    \centering
    \includegraphics[width=0.9\linewidth]{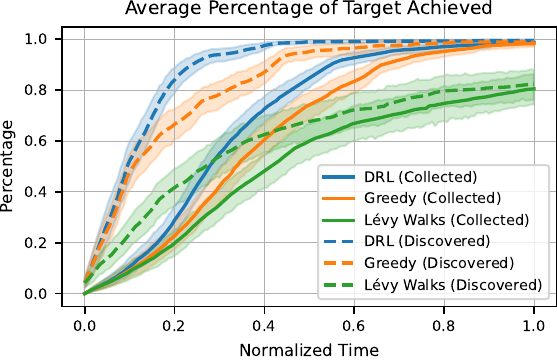}
    \caption{Percentage of Target Achieved (PTA) metric for the objective of each team over normalized time, averaged across 100 episodes for 3 algorithms. Confidence intervals are represented as shaded regions.}
    \label{fig:A.a_average_pta_both_teams_comb_port}
\end{figure}

The evaluation scenario selected to carry out the comparison is based in a typical sport wharf, showed in Figure~\ref{fig:food_render}, with narrow corridors and difficult access between points, presented in \cite{mendozab}. This scenario is formed by a matrix of $H=62 \times W=46$, with $1297$ visitable nodes.
At each episode a unique food distribution is generated. A initial hotspot is randomly selected from all visitable nodes, and around it items are randomly scattered according a multivariate normal distribution, ensuring uniform distribution of the elements from an emission source.
The number of items is also drawn from a normal distribution at the beginning of each episode. Items are dynamic, that is, they move at each time step following two components. First, environmental currents, represented as a wind direction constant during the episode, but sampled from a uniform distribution at the beginning of it. Secondly, a random motion component is added to each item at each time step to increase the unpredictability of dynamics.
Each episode has a maximum duration of 150 steps, which may be shorter if agents manage to eliminate all elements before the time limit.
This upper bound, which leaves a reasonable margin for completing the mission, ensures a fair comparison between algorithms by standardizing the available interaction budget while capturing efficiency differences when missions end early.
Two agents for each team are deployed in the environment, resulting in a four heterogeneous multi-agent system composed of specialized roles. 
Maintaining a fixed team size allows a controlled evaluation, particularly when analyzing the effect of degrading the behavior by removing agents of one team.

Below it is presented a detailed analysis of all proposed evaluation metrics, illustrating their behavior across the different algorithmic configurations. Metrics are examined in terms of its interpretation, comparative results, and the insights it provides into coordination dynamics, division of labor, and efficiency within the heterogeneous multi-agent foraging environment.

\begin{table}[t]
\centering
\caption{Primary Metrics: Average Mean and confidence interval (CI) of PTA and RMSE at the end of episodes.}
\label{tab:metrics_pta_rmse}
\resizebox{\columnwidth}{!}{
\begin{tabular}{lcccccc}
\toprule
\multirow{2}{*}{\textbf{Algorithm}} &
\multicolumn{2}{c}{$\textbf{PTA}_\textbf{D}$ (\%)} &
\multicolumn{2}{c}{$\textbf{PTA}_\textbf{C}$ (\%)} &
\multicolumn{2}{c}{\textbf{RMSE}} \\
\cmidrule(lr){2-3} \cmidrule(lr){4-5} \cmidrule(lr){6-7}
& Mean & CI ($95\%$) & Mean & CI ($95\%$) & Mean & CI ($95\%$) \\
\midrule
DRL & 99.41 & 0.46 & 98.68 & 0.56 & 0.0010 & 0.0005 \\
Greedy & 98.85 & 1.04 & 98.20 & 1.15 & 0.0017 & 0.0012 \\
L\'evy Walks & 82.28 & 5.93 & 80.43 & 5.98 & 0.0165 & 0.0051 \\
\bottomrule
\end{tabular}
}
\end{table}

\subsection{Primary metrics}\label{Section: Primary metrics}
Figure~\ref{fig:A.a_average_pta_both_teams_comb_port} illustrates the evolution of the average PTA over time for the different evaluated algorithms and teams, with their confidence intervals. In this context, metric reflects different objectives for each team: for the scout team, it measures the percentage of resources (food items) discovered $PTA_D$, while for the forager team it represents the percentage of resources successfully collected $PTA_C$. 
The  joint interpretation of curves with Table~\ref{tab:metrics_pta_rmse}, which shows the final RMSE and PTA metrics with their confidence intervals, reveal clear differences in algorithms overall task efficiency, allowing for direct comparison of convergence speed and final performance levels.

DRL-based approach demonstrates a noticeably faster initial growth rate for both teams, indicating more effective early decision-making and exploration–exploitation balance. This can also be seen numerically in Table~\ref{tab:metrics_nt_throughput}, which shows the normalized time for the forager team to reach $NT_{50\%}$ and $NT_{90\%}$ of $PTA_C$ metric, i.e., to eliminate that percentage of the total elements in the scenario during the episode.
In contrast, heuristic strategies, such as Greedy and Lévy Walks, show slower initial progress. However, the Greedy policy experiences a notable increase in the middle of the episode, allowing it to temporarily approach the performance of the DRL-based method. This behavior indicates that, although Greedy approach lacks long-term planning, it is capable of effectively exploiting local opportunities when they appear densely clustered.
For its part, the Lévy Walks strategy falls significantly behind in terms of $PTA_{D}$ (ratio of discovered items), due to its highly random nature that does not take into account either the shape or the information gathered from the environment. However, despite being a more basic algorithm, most of the items that are detected end up being collected. 
Lévy Walks allows for broad global coverage in very open scenarios, taking advantage of its exploratory nature to travel long distances, but its ability to concentrate and exploit resource-rich regions is limited, especially when the dynamics of elements require informational and directional updating, since its movements are not adaptive.

\begin{table}[t]
\centering
\caption{Primary Metrics: Normalized time to reach target collection levels and final throughput.}
\footnotesize
\label{tab:metrics_nt_throughput}
\begin{tabular}{lccc}
\toprule
\textbf{Algorithm} & $\textbf{NT}_{\text{50\%}}^{\textbf{PTA}_\textbf{C}}$ & $\textbf{NT}_{\text{90\%}}^{\textbf{PTA}_\textbf{C}}$ & \textbf{Throughput} \\
\midrule
DRL & 0.281 & 0.512 & 0.245 \\
Greedy & 0.349 & 0.583 & 0.244 \\
L\'evy Walks & 0.477 & 0.735 & 0.199 \\
\bottomrule
\end{tabular}
\end{table}

\begin{figure}[t]
    \centering
    \includegraphics[width=0.9\linewidth]{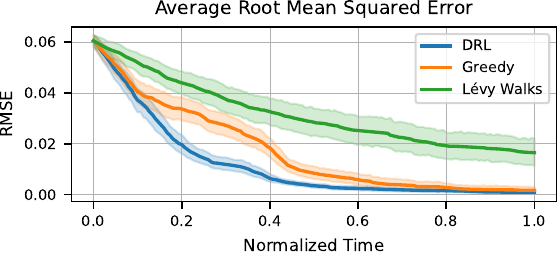}
    \caption{Root Mean Squared Error (RMSE) of the items distribution over normalized time, averaged across 100 episodes for 3 algorithm. Confidence intervals are represented as shaded regions.}
    \label{fig:A.b_average_rmse_comb_port}
\end{figure}

A similar trend to that observed in the PTA metric can be seen in the average RMSE of the estimated food distribution presented in Figure~\ref{fig:A.b_average_rmse_comb_port}. In this case, the RMSE reflects the accuracy of the spatial–temporal estimation of resources. Lower values indicate that agents are able to more reliably anticipate the evolution of the environment.
Consistently with the PTA results, the DRL-based method achieves the fastest and most stable reduction in prediction error, while Greedy exhibits a slower but noticeable improvement. Lévy Walks strategy maintains a significantly higher RMSE with a nearly constant slope, aligning with its exploratory nature that does not incorporate informed decision-making. 
This supports the observation that although Lévy enables broad coverage of the environment, it fails to refine knowledge of high–value areas.

\begin{figure}[t]
    \centering
    \includegraphics[width=0.9\linewidth]{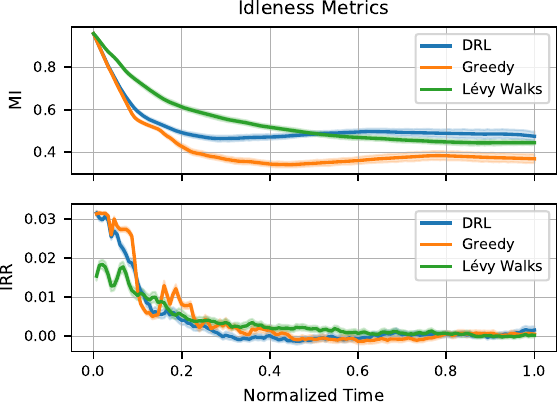}
    \caption{Idleness metrics over normalized time, averaged across 100 episodes for 3 algorithm. Confidence intervals are represented as shaded regions.}
    \label{fig:A.c_idleness_reduction_rate_comb_port}
\end{figure}

This interpretation is further supported by the MI and IRR metrics, shown in Figure~\ref{fig:A.c_idleness_reduction_rate_comb_port}, which measures how frequently each region of the map is revisited. In this case, lower values correspond to more balanced and persistent coverage of the environment.
Greedy algorithm exhibits the fastest and most sustained reduction in MI, which indicates that it continuously focuses its intensification on the least recently visited areas. This behaviour results from its purely myopic nature: agents always select the action that yields the highest immediate reward, without considering long–term benefits. As a result, Greedy tends to aggressively move toward locations that currently promise improvement in coverage, as seen in IRR, rather than positioning itself in regions that may contain richer resource densities in the future, but as seen in Figures~\ref{fig:A.a_average_pta_both_teams_comb_port} and ~\ref{fig:A.b_average_rmse_comb_port}, it is something that demonstrates strong effectiveness for this type of problem.

DRL initially shows a sharp decline in MI in Figure~\ref{fig:A.c_idleness_reduction_rate_comb_port}, reflecting an initial phase of rapid coverage of the environment, but subsequently stabilizes at higher levels than Greedy. Notably, it is eventually overtaken by the Lévy Walks strategy. This behavior seems to suggest that DRL scouts, after forming an global representation of the environment, shift their attention toward updating areas already known to contain resources rather than continuing broad exploration. For that reason, RMSE in Figure~\ref{fig:A.b_average_rmse_comb_port} continues to decrease, while MI stabilizes.
Conversely, Lévy Walks maintains moderate MI values with progressive improvement, consistent with its inherently broad and stochastic exploration pattern, which repeatedly sweeps large areas of the environment without intentional prioritization. 

Table~\ref{tab:metrics_nt_throughput} also shows Throughput, measured as the average number of items collected per step and agent. The obtained values reveal that DRL and Greedy achieve almost identical productivity, while Lévy remains noticeably lower. This result is consistent with the trends observed in the PTA and Idleness metrics: although Greedy lacks long–term strategic reasoning, its strongly exploitative behavior enables rapid conversion of nearby discoveries into processed items, sustaining a competitive collection rate comparable to DRL. In contrast, DRL achieves similar throughput through a more balanced distribution of effort, prioritizing informed movement based on accumulated knowledge, as reflected by its superior RMSE performance. Meanwhile, the lower Throughput exhibited by Lévy approach aligns with its exploratory nature—capable of wide coverage and discovery potential, but less effective in concentrating agents in high–value regions for resource removal. Primary metrics reinforces the idea that comparable performance may emerge from fundamentally different algorithms, highlighting the importance of analyzing multiple complementary metrics.

\subsection{Inter-team metrics evaluation}\label{Section: Inter-team metrics}
\begin{figure}[t]
    \centering
    \includegraphics[width=0.9\linewidth]{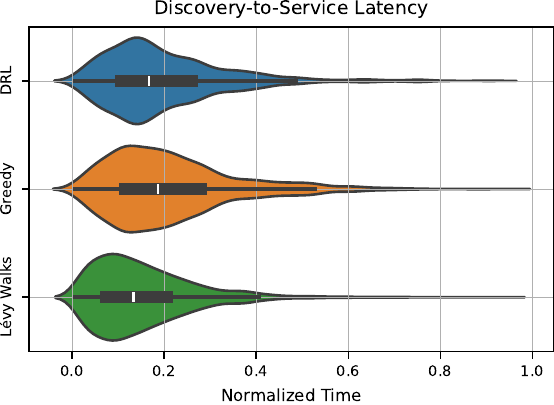}
    \caption{Distribution representation of Discovery-to-Service Latency (DSL) metric during 100 episodes for 3 algorithms.}
    \label{fig:B.a_discovery_to_service_latency}
\end{figure}

Figure~\ref{fig:B.a_discovery_to_service_latency} presents the \emph{Discovery-to-Service Latency} metric, measuring the normalized time between the moment a scout discovers an item and the moment a forager successfully processes it. The distribution representation reveals that the Lévy Walks strategy achieves the lowest average latency, followed by DRL and Greedy. This superior performance of Lévy can be attributed to the deterministic behavior of its foragers, which has been designed so that when an element is discovered, they follow the shortest Dijkstra path to the nearest known target, ensuring fast resource servicing once an item is detected. In contrast, DRL exhibits slightly higher latency, likely due to policy-driven decision making that balances exploitation with broader situational reasoning, occasionally delaying immediate servicing in favor of long–term gain. Finally, a detailed inspection of the violin shape reflects that the higher average latency of Greedy seems to be due to the presence of a significantly long tail.
This indicates that, although Greedy achieves fast discovery-to-service cycles, it also frequently experiences extreme delays. These long latency events are likely to occur when agents get stuck in locally optimal decisions and are unable to reach rapidly the items present in the scenario. Overall, these results demonstrate how latency reveals operational differences in coordination efficiency that are not apparent from individual performance metrics such as PTA or Throughput.

\begin{figure}[t]
    \centering
    \includegraphics[width=0.9\linewidth]{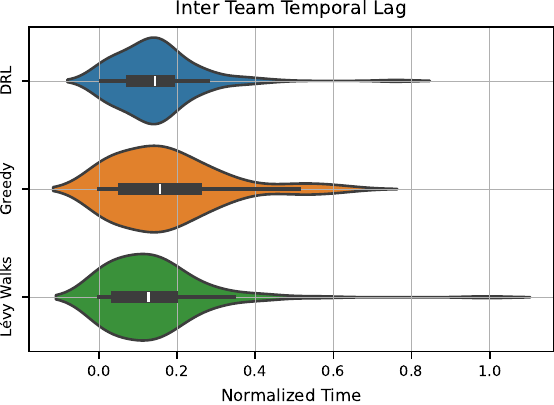}
    \caption{Distribution representation of Inter-Team Temporal Lag (ITL) metric during 100 episodes for 3 algorithms.}
    \label{fig:B.b_inter_team_temporal_lag}
\end{figure}

The distribution representation from Figure~\ref{fig:B.b_inter_team_temporal_lag} for the ITL metric reveals similar insights into the temporal coordination capabilities of the algorithms. 
It measures the temporal offset between the peak PTA achievement of scouts ($PTA_D$) and foragers ($PTA_C$), capturing how synchronized both teams are in completing their respective sub-objectives. 
The results show that the Greedy strategy again exhibits slightly worst performance.
In the case of DRL, the distribution is more concentrated near its mean, indicating a more consistent synchronization between the peak of discovery and the peak of servicing. This stability can be attributed to the strategic positioning of foragers relative to scouts: through learning, foragers implicitly exploit the expectation that resource discoveries are more likely to occur in their vicinity, reducing unnecessary travel and resulting in tightly clustered ITL values. In contrast, both Lévy and Greedy exhibit more rounded and dispersed violin shapes, reflecting greater variability in this metric. It is, therefore, a metric with results similar to DSL, but with a slightly different approach, as it focuses on performance peaks.

\begin{figure}[t]
    \centering
    \includegraphics[width=0.9\linewidth]{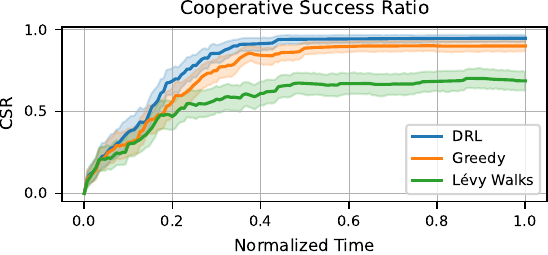}
    \caption{Cooperative Success Ratio (CSR) metric over normalized time, averaged across 100 episodes for 3 algorithm. Confidence intervals are represented as shaded regions.}
    \label{fig:B.c_cooperative_success_ratio}
\end{figure}

Another informative metric is the \textit{Cooperative Success Ratio}, which quantifies the fraction of tasks completed through explicit cooperation between the two teams, that is, the proportion of items that were first discovered by scouts and subsequently processed by foragers. Figure~\ref{fig:B.c_cooperative_success_ratio} shows the temporal evolution of this metric, while Table~\ref{tab:metrics_csr_ss} reports the final values. Initially, all three algorithms display similar cooperative performance, but around \(20\%\) of the episode, DRL and Greedy experience a noticeable increase, with a similar final value. This can be explained by the operational dependencies designed in the algorithms: foragers in both DRL and Greedy are exclusively focused on servicing known items, making them highly reliant on scouts for discovery. Consequently, as scouts discover information about food location, the cooperation ratio naturally grows. In contrast, Lévy Walks foragers also perform exploratory movements when the known item model is empty, following the stochastic Lévy patterns. While this allows for broader environmental coverage, it reduces the strict dependence on scouts, leading to lower cooperative success overall. The analysis of this metric reflects how design choices directly influence the emergence and efficiency of inter-team cooperation.

\begin{figure}[t]
    \centering
    \includegraphics[width=1\linewidth]{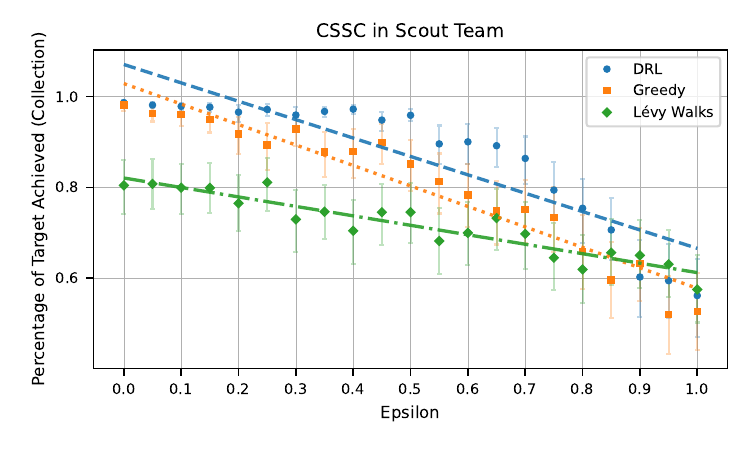}
    \caption{Cooperation Sensitivity under Stochastic Corruption ($CSSC$) for the scout team evaluated on the collection performance metric ($PTA_C$). The best-fit line highlights the rate at which collection performance degrades with each algorithm. Confidence interval 95\% is included.}
    \label{fig:B.e_epsilon_study_ptaForagers_EpsTmScout_comb_port_pointplot}
\end{figure}

The \textit{Cooperation Sensitivity under Stochastic Corruption} analysis evaluates how the performance of one team is affected when the behavior of the other team is progressively degraded. To do so, an increasing proportion of random actions is injected into the scout policy, controlled by a parameter \(\varepsilon \in [0,1]\) in steps of \(0.05\), while foragers follow their normal decision policy, and then, in the opposite way. For a fair comparison, the random actions are generated using the same seed across all algorithms. Figure~\ref{fig:B.e_epsilon_study_ptaForagers_EpsTmScout_comb_port_pointplot} shows the evolution of the $PTA_C$ metric at the end of evaluation episodes as \(\varepsilon_{scouts}\) increases. As expected, all algorithms exhibit performance deterioration as the scouts become less reliable. However, the slope ($SS$) of the best–fit line reveals different levels of sensitivity, as numerically shown in Table~\ref{tab:metrics_csr_ss}.

This behaviour can be interpreted through algorithmic characteristics. For Lévy, the lower sensitivity is partially explained by its inherently lower baseline performance—starting from a PTA significantly below than the nearly maximum achieved by DRL and Greedy, the margin for degradation is smaller. Additionally, the exploration strategy of Lévy already relies heavily on stochastic movement patterns, meaning that adding further randomness produces a comparatively mild perturbation. In contrast, both DRL and Greedy depend strongly on scouts: their foragers allocate effort purely to servicing known items, making them more vulnerable to corrupted scout behavior. Between these two, DRL shows noticeably greater robustness, reflected in its lower absolute $SS_{scouts}^{PTA_C}$. This can be attributed to the ability of the learned policy to generalize under noisy conditions, whereas Greedy strictly maximizes immediate reward without any anticipatory structure, collapsing more abruptly under stochastic corruption.

\begin{figure}[t]
    \centering
    \includegraphics[width=1\linewidth]{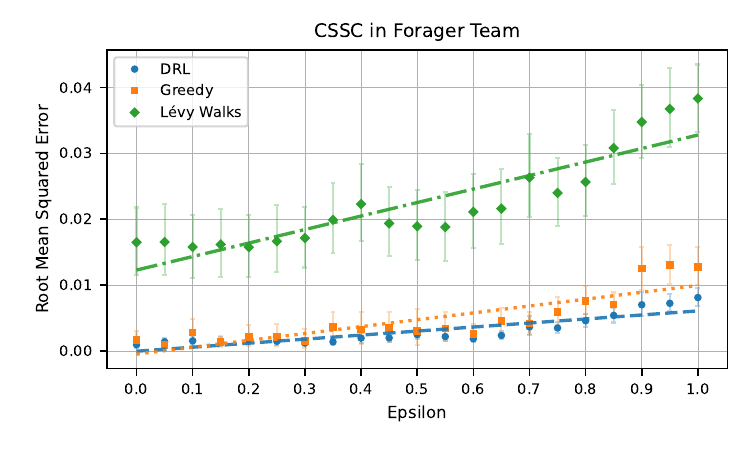}
    \caption{Cooperation Sensitivity under Stochastic Corruption ($CSSC$) for the forager team evaluated on the model RMSE. The best-fit line highlights the rate at which model degrades with each algorithm. Confidence interval 95\% is included.}
    \label{fig:B.e_epsilon_study_RMSE_EpsTmForager_comb_port_pointplot}
\end{figure}

Figure~\ref{fig:B.e_epsilon_study_RMSE_EpsTmForager_comb_port_pointplot} reports how the predicted model error (RMSE) evolves when stochastic corruption with parameter \(\varepsilon_{foragers}\) increases. Across all algorithms the RMSE increases as epsilon grows, but the magnitude and rate of degradation differ: Lévy exhibits the largest deterioration, Greedy shows a slightly more pronounced than the DRL slope, while DRL is the most robust to the noise introduced.
Several interrelated causes explain this pattern. When foragers are corrupted and fail to remove discovered resources, two effects combine to raise the RMSE maintained by the scouts. First, detected items remain in the environment longer and continue to move, so the scouts' previous observations rapidly become outdated; the predictive model therefore accumulates systematic error. Second, missed removals increase the number of elements in the environment, so it is harder to keep the position of all of them up to date, degrading the quality of the model. Both effects are stronger the longer the discovery-to-service loop is broken, which is precisely what happens as \(\varepsilon\) increases.
Algorithm-specific factors also modulate this sensitivity. DRL scouts (and DRL algorithm as a whole) learn behaviors about the spatial distribution and dynamics of food, enabling more informative updates and implicit recovery strategies. This learned adaptivity translates into a smaller ${SS_{forag}^{RMSE}}$ under forager team corruption. Greedy scouts, although driven by local information and immediate reward, still incorporate deterministic rules attached to observed item positions and thus degrade faster than DRL but more gradually than Lévy. In contrast, Lévy scouts follow essentially stochastic exploration trajectories and do not prioritize model-driven updates; hence, when foragers fail to remove items the scouts neither compensate by re-sampling high-value locations, producing the largest RMSE increase.
Furthermore, from previous metrics it had also been deduced that Lévy leaves part of exploration tasks in the hands of the foragers. The more a system relies on foragers to maintain environmental truth, the more the predictive model degrades when those foragers are corrupted.

Finally, as mentioned in the definition of the metric, Figures~\ref{fig:B.e_epsilon_study_ptaForagers_EpsTmScout_comb_port_pointplot} and~\ref{fig:B.e_epsilon_study_RMSE_EpsTmForager_comb_port_pointplot} show that the degradation curves have distinct sections in which different levels of degradation are perceived when noise is introduced. In particular, the DRL-based approach shows a two-part behavior, especially in Figure~\ref{fig:B.e_epsilon_study_ptaForagers_EpsTmScout_comb_port_pointplot}, with an initial region in which performance degrades slowly as $\epsilon$ increases, followed by a more pronounced drop beyond a certain corruption threshold, around $\epsilon=0.5$. This indicates the existence of two distinct sensitivity regimes: an initial robust regime in which the learned policy tolerates moderate stochastic perturbations, and a second regime in which cooperation suddenly deteriorates once noise exceeds the learned coordination patterns. Greedy, for example, shows a more gradual decline. This behavior is particularly relevant in practical implementations, as it suggests that DRL-based cooperation is more resilient than Greedy under mild policy degradation (e.g., communication noise, model inaccuracies, or partial failures), even if both eventually collapse under severe corruption.

\begin{table}[t]
\centering
\caption{Inter-Team Metrics: Final Average Cooperative Success Ration and Sensitivity Slope Metrics.}
\footnotesize
\label{tab:metrics_csr_ss}
\begin{tabular}{lccc}
\toprule
\textbf{Algorithm} & $\mathbf{CSR}$ & $\mathbf{SS_{scouts}^{PTA_C}}$ & $\mathbf{SS_{forag}^{RMSE}}$ \\
\midrule
DRL & 0.946 & -40.55 & 0.0061 \\
Greedy & 0.899 & -45.14 & 0.0104 \\
L\'evy Walks & 0.686 & -20.86 & 0.0205 \\
\bottomrule
\end{tabular}
\end{table}

\subsection{Intra-team metrics evaluation}

\begin{figure}[t]
    \centering
    \includegraphics[width=1\linewidth]{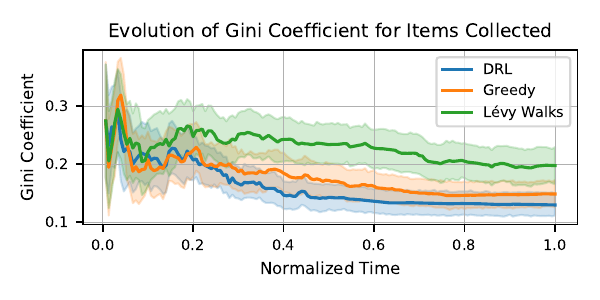}
    \caption{Gini Coefficient metric in forager team over normalized time, averaged across 100 episodes for 3 algorithm. Confidence intervals are represented as shaded regions.}
    \label{fig:C.a_gini_coefficient_items_collection_comb_port}
\end{figure}

Figure~\ref{fig:C.a_gini_coefficient_items_collection_comb_port} shows the temporal average evolution of the Gini Coefficient applied to the workload of the foragers within each team. Given that there are only two foragers, the maximum possible value of the coefficient would be 0.5 (one works completely while the other remains inactive), and the minimum 0 (perfectly balanced work). At the beginning of the episode, all algorithms present similar values, around 0.2–0.3, reflecting a moderately unequal but still unstable distribution due to the initial phase of exploration and irregular discovery of elements. However, approximately between 20\% of the episode's length, a clear divergence is observed: both DRL and Greedy begin to consistently reduce their internal inequality, stabilizing around 60\% of the episode at values clearly lower than the initial ones. The final balance is more pronounced in DRL, indicating better operational coordination and efficient distribution of work among its foragers. In contrast, in Lévy Walks, the curve shows a significantly smoother, practically flat decline, implying a persistence of inequality in the service load among agents.
What is more interesting about this result is that the greater equity observed in DRL and Greedy is not due to an explicit imposed  cooperative distribution mechanism, but rather to an emergent effect of how they operate or position themselves on the map. 
This effect is especially noticeable in DRL, where the learned policy incorporates information about the state of the environment and the positions of other agents, suggesting that they learn to select actions that tend to avoid interference and redundancy, as their reward would be lower. Despite this quality of the DRL, it does not seem to be the only reason for its good equity in the Gini Coefficient.
Greedy, despite lacking awareness of teammates and optimizing only for immediate local reward, also achieves a relatively balanced workload. This balance suggest that it does not arise from cooperative reasoning, but from properties induced by the speed at which targets become available: as shown above in Section~\ref{Section: Primary metrics} with the evolution of PTA metric, both DRL and Greedy show rapid and continuous discovery of new elements, leading to frequent incorporation of targets into the environment model. Under these conditions, a single collector cannot exploit all discovered resources on its own, so both agents inevitably capture tasks that would otherwise be accumulated by only one, leading to emergent fairness.
In contrast, Lévy Walks produce a slower and more irregular pattern of discovery, in which resources appear in smaller batches and at longer intervals. When only a few targets are known at a time, the collector located closest to them tends to monopolize the available tasks, while the other collector arrives at the location too late. As a result, the imbalance in workload persists for much longer, which is reflected in significantly higher Gini values.

\begin{figure}[t]
    \centering
    \includegraphics[width=0.9\linewidth]{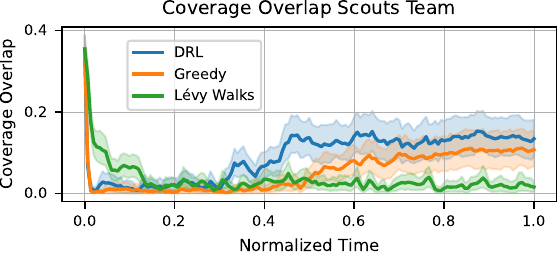}
    \caption{Coverage Overlap (CO) metric in scout team over normalized time, averaged across 100 episodes for 3 algorithm. Confidence intervals are represented as shaded regions.}
    \label{fig:C.b_coverage_overlap_comb_port}
\end{figure}

Figure~\ref{fig:C.b_coverage_overlap_comb_port} shows the evolution of the \emph{Coverage Overlap} metric for the scout team, which reveals distinct behavioral characteristics across algorithms. As expected, all methods exhibit an initial peak, since all agents start from the same region and therefore observe overlapping areas. Shortly after the episode begins, overlap drops sharply: DRL and Greedy reach near-zero values around 10\% of the episode duration, while Lévy Walks takes a while longer to fall. Lévy shows a longer initial overlap because its agents, following pseudo-random trajectories with many short steps, take longer to diverge spatially, unlike DRL and Greedy, which introduce more directional decisions that force a faster separation.
Once separated, Lévy remains with almost zero overlap for the remainder of the episode, with an extremely narrow confidence band. This reflects its highly exploratory and stochastic nature, where agents continue dispersing globally, covering largely dispersed regions.

In contrast, both DRL and Greedy show a progressive increase after the initial drop, stabilizing around 0.15 and 0.10 respectively. This suggests that, although both strategies prioritize coverage, they tend to converge in regions where valuable clusters of resources appear. In Greedy, this behavior arises naturally: when a high-density area is discovered, both collectors are attracted to the same targets due to the reward for approaching the nearest resource. In DRL, a similar effect emerges, although in this case agents learn that, even when the reward in overlapping regions is shared, convergence toward high-value areas remains beneficial because it improves the accuracy of the shared model, given by the design of the reward function itself. Since DRL optimizes long-term rewards, scouts strategically reposition themselves around areas with high expected returns and where new discoveries are more likely to generate additional rewards. A moderate degree of overlap may be beneficial, enabling scouts to reinforce local updates of dynamically changing resource locations, but excessive overlap would waste coverage.

As previously analyzed in Section~\ref{Section: Primary metrics}, Table~\ref{tab:metrics_pta_rmse} reports the absolute performance achieved by the complete system: two scouts and two foragers, establishing a common reference for all algorithms. DRL and Greedy achieve near-perfect performance both in items discovery and removal, with extremely low baseline RMSE values on the order of $10^{-3}$, whereas Lévy Walks exhibits significantly higher model error and lower task completion. Table~\ref{tab:ablation_results} presents the corresponding ablation setups used to compute the Marginal Contributions metric, where either one scout or one forager is removed. 

In all algorithms, removing an scout appears to have a more detrimental impact on performance metrics than removing a forager. This indicates that exploration and information gathering are the most critical factor for both the execution of collection tasks and model accuracy in the proposed scenario. Even in methods such as Lévy Walks, which employs foragers to explore when there are no known items, a decrease in dedicated scouts leads to a significant reduction in the percentage of items discovered, ultimately limiting the effectiveness of subsequent collection actions. These findings suggest that, despite their indirect role in collection execution, scouts act as a fundamental facilitator for agent coordination.
A similar effect is observed in the RMSE, which increases significantly more when eliminating a scout than when eliminating a forager.
In DRL and Greedy it only increases moderately when a forager is removed, but this increase is mainly attributed to the greater number of active elements remaining in the environment, rather than to a deterioration in exploration efficiency, since the discovery rate ($PTA_{D}$) remains almost unchanged. In contrast, Lévy Walks relies on foragers for exploration when no targets are known, which causes $PTA_{D}$ to decrease more noticeably than in the other methods when a forager is removed.
These patterns of behavior are consistent with previous deductions based metrics studied in Sections~\ref{Section: Primary metrics} and~\ref{Section: Inter-team metrics}.

\begin{table}[t]
\centering
\caption{Inter-Team Metrics: Performance under ablated team configurations used for Marginal Contribution analysis versus complete setup. 
}
\label{tab:ablation_results}
\resizebox{\columnwidth}{!}{
\begin{tabular}{lcccc}
\toprule
\textbf{Algorithm} & \textbf{Setup} & $\mathbf{{PTA}_{D}}$ \textbf{(\%)} & $\mathbf{{PTA}_{C}}$ \textbf{(\%)} & $\mathbf{RMSE}$ \\
\midrule
\multirow{3}{*}{DRL} & Complete & 99.41 & 98.68 & 0.0010 \\
 & 1S--2F & 87.92 & 93.55 & 0.0077 \\    & 2S--1F & 95.14 & 99.32 & 0.0017 \\
\midrule
\multirow{3}{*}{Greedy} & Complete & 98.85 & 98.20 & 0.0017 \\
 & 1S--2F & 82.26 & 88.08 & 0.0130 \\
 & 2S--1F & 94.89 & 98.70 & 0.0029 \\
\midrule
\multirow{3}{*}{Lévy Walks} & Complete & 82.28 & 80.43 & 0.0165 \\
 & 1S--2F & 72.50 & 76.17 & 0.0223 \\
 & 2S--1F & 73.85 & 78.60 & 0.0202 \\
\bottomrule
\end{tabular}
}
\end{table}

\begin{table}[t]
\centering
\caption{Inter-Team Metrics: Role-based Marginal Contribution (MC).}
\scriptsize
\setlength{\tabcolsep}{4.3pt} 
\label{tab:metrics_mc}
\begin{tabular}{lcccccc}
\toprule
\multirow{2}{*}{\textbf{Algorithm}} & \multicolumn{3}{c}{\textbf{MC (1 Scout)}} & \multicolumn{3}{c}{\textbf{MC (1 Forager)}} \\
\cmidrule(lr){2-4} \cmidrule(lr){5-7}
 & $\mathbf{PTA_D}$ & $\mathbf{PTA_C}$ & $\mathbf{RMSE}$ & $\mathbf{PTA_D}$ & $\mathbf{PTA_C}$ & $\mathbf{RMSE}$ \\
\midrule
DRL & 0.0589 & 0.1090 & 0.8701 & 0.0009 & 0.0359 & 0.4118 \\
Greedy & 0.1090 & 0.1623 & 0.8692 & 0.0015 & 0.0337 & 0.4138 \\
L\'evy Walks & 0.0743 & 0.0986 & 0.2601 & 0.0447 & 0.0818 & 0.1832 \\
\bottomrule
\end{tabular}
\end{table}

\emph{Marginal Contribution} reported in Table~\ref{tab:metrics_mc} quantifies the individual impact of the number of agents in a team in a standardized form, but it is important to contextualize the information with the values in Table~\ref{tab:ablation_results}.
MC metric reveals that DRL and Greedy exhibit very high marginal contributions of the scout agents with respect to RMSE. However, this effect is largely driven by scale: since both algorithms operate with extremely low baseline errors in the complete system (as seen in Table~\ref{tab:metrics_pta_rmse}), even small absolute degradations in the ablated settings (Table~\ref{tab:ablation_results}) translate into large relative contributions. In contrast, Lévy Walks operates with a much higher reference RMSE, so the marginal contribution of a scout agent in RMSE is lower.
Moreover, configurations with fewer foragers lead to higher food accumulation, which further increases the difficulty of maintaining an accurate model independently of the cooperative strategy, as seen in MC metric. 

The marginal contributions associated with $PTA_D$ and $PTA_C$ further clarify the asymmetric roles of both teams. For $PTA_D$, removing a scout produces a consistent and significant degradation across all algorithms, with particularly strong effects in Greedy and DRL, confirming the central role of scouts in information acquisition. In contrast, removing a forager yields a nearly negligible MC in $PTA_D$ for DRL and Greedy, indicating that foragers play a limited role in the discovery process in these methods. Lévy Walks exhibits a different pattern: the MC of a forager in $PTA_D$ becomes non-negligible, reflecting its reliance on foragers for exploratory behavior in the absence of known targets.

For the collection metric $PTA_C$, the marginal contribution analysis reveals that removing a scout produces a larger degradation than removing a forager across all algorithms. This indicates that, although foragers are the agents directly responsible for collecting, the availability of accurate and timely information provided by scouts is a stronger limiting factor for effective collection. In DRL and Greedy, this effect is particularly pronounced, reflecting their strong dependence on scout-driven discovery to sustain efficient collecting policies. 
In Lévy Walks, a remarkably similar level of degradation is observed in $PTA_C$ when either a scout or a forager is removed, due to the qualities of the algorithms mentioned above.

\section{Conclusions}

In this work, it has been addressed the problem of characterizing and quantifying sequential cooperation in heterogeneous multi-agent systems operating under partial observability and dynamic task distributions.  
To this end, we have developed, formalized, and experimentally analyzed a comprehensive set of cooperation metrics tailored to the interaction between exploration and servicing teams in sequential foraging environments.
The proposed metrics are organized into three major groups: primary performance metrics, inter-team cooperation metrics, and intra-team metrics.
Results suggest that classical performance indicators alone are insufficient to fully describe cooperative behavior, as isolated metrics can lead to partial or even misleading interpretations.
The experimental study confirms that the proposed metrics enable a significantly richer and more interpretable analysis of cooperation dynamics when analyses jointly, offering a solid foundation for future research on cooperative mechanisms.
They can serve as practical tools to guide the design of new algorithms, as they explicit reveal coordination structures, sensitivities, and inefficiencies that are not visible through 
commonly used approaches.

Besides, several limitations must be acknowledged. The study focuses on a specific dual role structure (scouts and foragers), which may restrict the direct generalization of some metrics to more complex team hierarchies.
However, it is worth noting that the simplifying assumptions adopted in this work, such as agent mobility, fixed and initially unknown resource distributions and no regeneration, unlimited carrying capacity, or environment topology, do not limit the generality or validity of the proposed metrics. 
The metrics are independent to these design choices and can be helpful in designing others more cooperative scenarios.

Future work could extend this framework in several directions, such as its implementation in more realistic foraging platforms, incorporating explicit energy constraints, communication latency, packet losses, and limited bandwidth, which would introduce new restrictions. 
Additionally, the framework could be applied to scenarios with more than two functional roles and team organizations.

\bibliographystyle{elsarticle-num}

\bibliography{bib}

\end{document}